\pdfoutput=1
\documentclass[paper=a4]{scrartcl}

\usepackage[utf8]{inputenc}
\usepackage[T1]{fontenc}
\usepackage{amsfonts}
\usepackage{amsmath}
\usepackage{amssymb}
\usepackage{lmodern}
\usepackage{enumerate}
\usepackage[version=4]{mhchem}
\usepackage{hyperref}

\newcommand\numberthis{\addtocounter{equation}{1}\tag{\theequation}}
\newcommand{\dotequal}{\mathrel{\dot{=}}}

\newcommand{\QQ}{\mathbb{Q}}
\newcommand{\RR}{\mathbb{R}}
\newcommand{\ZZ}{\mathbb{Z}}

\newcommand{\ff}{\mathbf{f}}
\newcommand{\gggg}{\mathbf{g}}

\newcommand{\kk}{\mathbf{k}}
\newcommand{\tttt}{\mathbf{t}}
\newcommand{\vv}{\mathbf{v}}
\newcommand{\xx}{\mathbf{x}}
\newcommand{\yy}{\mathbf{y}}

\newcommand{\Vk}{V_\kk}
\newcommand{\Vks}{V_{\kk^*}}
\newcommand{\Rp}[1]{\RR_{>0}^{#1}}
\newcommand{\Rs}[1]{{\RR^*}^{#1}}
\newcommand{\Rsn}{\Rs{n}}
\newcommand{\false}{\operatorname{false}}
\newcommand{\true}{\operatorname{true}}

\newcommand{\PHI}[1]{\varphi_{#1}}
\newcommand{\PHIP}[1]{\PHI#1'}
\newcommand{\COSET}[1]{\varphi_{\text{coset}}}
\newcommand{\GROUP}[1]{\varphi_{\text{group}}}
\newcommand{\kf}[1]{k^{\text{f}}_{#1}}
\newcommand{\kr}[1]{k^{\text{r}}_{#1}}
\newcommand{\kon}[1]{k^{\text{on}}_{#1}}
\newcommand{\koff}[1]{k^{\text{off}}_{#1}}
\newcommand{\kcat}[1]{k^{\text{cat}}_{#1}}
\newcommand{\lon}[1]{\ell^{\text{on}}_{#1}}
\newcommand{\loff}[1]{\ell^{\text{off}}_{#1}}
\newcommand{\lcat}[1]{\ell^{\text{cat}}_{#1}}

\allowdisplaybreaks

\KOMAoptions{
  abstract=true,
  fontsize=10pt,
  captions=tableheading,
  headinclude=false,
  footinclude=false}

\addtokomafont{author}{\large}

\recalctypearea

\begin{document}
  \title{Parametric Toricity of Steady State Varieties of Reaction
  Networks}

\author{Hamid Rahkooy\\
  MPI Informatics, Germany\\
  \texttt{hamid.rahkooy@mpi-inf.mpg.de}
  \and
  Thomas Sturm\\
  CNRS, Inria, and the University of Lorraine, France\\
  MPI Informatics and Saarland University, Germany\\
  \texttt{thomas.sturm@loria.fr}
}

\date{July 2021}

\maketitle

\begin{abstract}
  We study real steady state varieties of the dynamics of chemical reaction
  networks. The dynamics are derived using mass action kinetics with parametric
  reaction rates. The models studied are not inherently parametric in nature.
  Rather, our interest in parameters is motivated by parameter uncertainty, as
  reaction rates are typically either measured with limited precision or
  estimated. We aim at detecting toricity and shifted toricity, using a
  framework that has been recently introduced and studied for the non-parametric
  case over both the real and the complex numbers. While toricity requires that
  the variety specifies a subgroup of the direct power of the multiplicative
  group of the underlying field, shifted toricity requires only a coset. In the
  non-parametric case these requirements establish real decision problems. In
  the presence of parameters we must go further and derive necessary and
  sufficient conditions in the parameters for toricity or shifted toricity to
  hold. Technically, we use real quantifier elimination methods. Our
  computations on biological networks here once more confirm shifted toricity as
  a relevant concept, while toricity holds only for degenerate parameter
  choices.
\end{abstract}
  
\section{Introduction}
We study the kinetics of reaction networks in the sense of \emph{Chemical
  Reaction Network Theory} \cite{feinberg-book}. This covers also biological
networks that are not reaction networks in a strict sense, e.g., epidemic models
and signaling networks. The kinetics of reaction networks is given by ordinary
differential equations (ODE) $\dot\xx = \ff$ with polynomial vector field
$\ff \in \ZZ[\kk,\xx]$, where $\kk$ are positive scalar reaction rates and $\xx$ are
concentrations of species over time. Such ODE are typically derived using mass
action kinetics \cite[Sect. 2.1.2]{feinberg-book}. For fixed choices
$\kk^* \in \Rp{s}$, the real variety
$\Vks(\ff) = \{\,\xx^* \in \RR^n \mid \ff(\kk^*,\xx^*) = 0\,\}$ describes the set of
steady states of the system.

\sloppypar One famous example is the Michaelis--Menten network
\cite{MichaelisMenten:23r}, which describes an enzymatic reaction as follows:
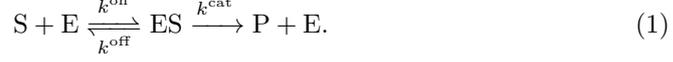
\begin{equation}\label{eq:mmnet}
  \ce{S + E <=>[$\kon{}$][$\koff{}$] ES ->[$\kcat{}$] P + E}.
\end{equation}
Here one has reaction rates $\kk = (\kon{}, \koff{}, \kcat{})$ and species
concentrations $\xx = (x_1, \dots, x_4)$ for the substrate $\ce{S}$, the enzyme
$\ce{E}$, the enzyme-substrate complex $\ce{ES}$, and the product $\ce{P}$,
respectively. The vector field of the ODE is given by $\ff = (f_1, \dots, f_4)$ as
follows, where $f_2=-f_3$:
\begin{align*}
  f_1 & = -\kon{}x_1x_2 + \koff{} x_3\\
  f_2 & = -\kon{}x_1x_2 + (\koff{} + \kcat{})x_3\\
  f_3 & = \kon{}x_1x_2 - (\koff{} + \kcat{})x_3\\
  f_4 & = \kcat{}x_3.\numberthis\label{eq:mmkin}
\end{align*}
For an intuition about mass action kinetics consider the reaction
$\ce{S + E ->[$\kon{}$] ES}$ in \eqref{eq:mmnet}. The summand $-\kon{}x_1x_2$ in
the differential equation $\dot{x_1} = f_1 = -\kon{}x_1x_2 + \koff{} x_3$
describes a decrease of the concentration $x_1$ of $\ce{S}$ that is proportional
to the product $x_1x_2$ of concentrations of $\ce{S}$ and $\ce{E}$ with a
positive proportionality factor $\kon{}$. The product $x_1x_2$ of concentrations
models the probability that one molecule of $\ce{S}$ and one molecule of
$\ce{E}$ encounter each other in a perfectly stirred reactor.

For steady state of the Michaelis--Menten kinetics, $f_4$ in \eqref{eq:mmkin}
imposes $x_3=0$. Biologically speaking, steady state requires that the
concentration of the enzyme-substrate complex become zero. Next, $f_1$, \dots,
$f_3$ impose that either $x_1=0$ and $x_2$ can be freely chosen, or vice versa.
That is, the concentration of either substrate or enzyme must become zero. The
concentration $x_4$ of the product can always be freely chosen. It turns out
that $\Vk(\ff) \neq \emptyset$, and $\Vk(\ff)$ does not depend on $\kk$ at all.

Let us look at 1-site phosphorylation
\cite{WangSontag:08p,Perez-MillanDickenstein:12a}, which gives a slightly more
complex network as follows:
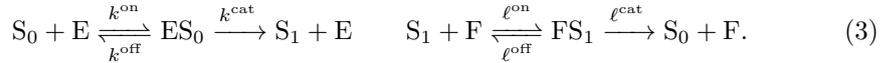
\begin{equation}\label{eq:pho1net}
  \ce{S_0 + E <=>[$\kon{}$][$\koff{}$] ES_0 ->[$\kcat{}$] S_1 + E}\quad\quad
  \ce{S_1 + F <=>[$\lon{}$][$\loff{}$] FS_1 ->[$\lcat{}$] S_0 + F}.
\end{equation}
Here we have $\kk = (\kon{}, \dots, \lcat{})$, $\xx = (x_1, \dots, x_6)$ for
concentrations of species $\ce{S_0}$, $\ce{S_1}$, $\ce{ES_0}$, $\ce{FS_1}$,
$\ce{E}$, $\ce{F}$, respectively. The vector field of the ODE is given by
$\ff = (f_1, \dots, f_6)$ with
\begin{align*}
  f_1 & =  - \kon{} x_1x_5 + \koff{} x_3 + \lcat{}x_4\\
  f_3 & = \kon{}x_1x_5 - (\kcat{} + \koff{})x_3\\
  f_4 & = \lon{}x_2x_6 - (\lcat{} + \loff{})x_4.\numberthis\label{eq:pho1kin}
\end{align*}
Similarly to $f_2$ in \eqref{eq:mmkin}, $f_2$, $f_5$, $f_6$ here are linear
combinations of $\ff'=(f_1,f_3,f_4)$ and thus $\Vk(\ff)=\Vk(\ff')$. In contrast
to the Michaelis--Menten kinetics we now find steady states where all species
concentrations are non-zero. One such steady state is
\begin{equation}\label{eq:pho1ss}
  \xx^* = \left(
  1, 1, 1, \frac{\kcat{}}{\lcat{}}, \frac{\kcat{} + \koff{}}{\kon{}},
  \frac{\kcat{}\lcat{} + \kcat{}\loff{}}{\lcat{}\lon{}}
  \right)^T.
\end{equation}
Notice that this particular steady state exists uniformly in $\kk$ and that
denominators cannot vanish, due to our requirement that $\kk > 0$.

For the non-parametric case, i.e., for fixed $\kk^* \in \Rp{s}$, comprehensive
computational experiments on reaction networks in \cite{GrigorievIosif:19a} have
identified \emph{shifted toricity} as a structural property that occurs
frequently but not generally. Assuming that $\Vks(\ff)$ is irreducible, the set
$\Vks(\ff)^* = \Vks(\ff) \cap \Rsn$ is \emph{shifted toric} if it forms a
multiplicative coset of $\Rsn$ \cite{grigoriev_milman2012}. Here $\Rs{}$ is the
multiplicative group of the field of real numbers, and $\Rsn$ is its direct
power. For the sake of this clear and simple algebraic setting, we do not take
into consideration the positivity of $\xx$ here. Instead, shifted tori can be
algorithmically intersected with the positive first orthant later on.

The notion of shifted toricity historically originates from the consideration of
additive groups. In our setting, the ``shift'' is geometrically not a
translation but a scaling of the torus. For the natural sciences, structural
properties like shifted toricity provide \emph{qualitative} insights into
nature, as opposed to quantitative information like numeric values of
coordinates of some fixed points. For symbolic computation, our hope is that
structural properties can be exploited for the development of more efficient
algorithms.

Our program for this article is the generalization of the concept of shifted
toricity to the parametric case, along with the development of suitable
computational methods, accompanied by computational experiments. For instance,
for our 1-site phosphorylation we will automatically derive in
Sect.\ref{se:npho} that
\begin{enumerate}[(i)]
\item\label{it:npho1i} $\Vk(\ff)^*$ forms a coset for all admissible choices of
  $\kk$, and
\item\label{it:npho1ii} $\Vk(\ff)^*$ forms a group if and only if
  $\kon{} - \koff{} = \lon{} - \loff{} = \kcat{} = \lcat{}$.
\end{enumerate}

Chemical reaction network theory \cite{feinberg-book} generally studies specific
structural properties of networks like \eqref{eq:mmnet} and \eqref{eq:pho1net},
such as our shifted toricity. There is a consensus in chemical reaction network
theory that meaningful structural properties of networks would not refer to
specific values of the rate constants $\kk$, as Feinberg explicitly states in
his excellent textbook: \textit{The network itself will be our object of study,
  not the network endowed with a particular set of rate constants}
\cite[p.19]{feinberg-book}. In reality, exact rate constants are hardly ever
known. They are either measured in experiments with a certain finite precision,
or they are estimated, often only in terms of orders of magnitude. Furthermore,
even if we had perfect positive real values for the rate constants $\kk$, recall
that according to mass action kinetics their co-factors are products of certain
species concentrations $\xx$, which only approximate probabilities as they would
hold under hypothetical ideal conditions. Hence, we are looking primarily for
results like (\ref{it:npho1i}) above. Result (\ref{it:npho1ii}) might seem
appealing from a mathematical viewpoint, but it has hardly any relevance in
nature. Bluntly speaking, a metabolism whose functioning depends on any of the
equations in (\ref{it:npho1ii}) could not be evolutionarily successful.

What is the motivation for looking at admissible parameter values at all? Why
not just derive yes/no decisions under suitable existential or universal
quantification of the parameters? First, just as the equations in
(\ref{it:npho1ii}) hardly ever hold in reality, the same arguments support the
hypothesis that derived inequalities, in the sense of logically negated
equations, in $\kk$ would hardly ever fail and may thus be acceptable. Second,
we are working in real algebra here. Even if there are no order inequalities in
the input, they will in general be introduced by the computation. For instance,
when asking whether there exists $x_1 \in \RR$ such that
$x_1^2 = k_1 - 10^6k_2$, an equivalent condition is given by $k_1 \geq 10^6k_2$.
Such a condition that one reaction rate be larger than another by several orders
of magnitude is meaningful and might provide useful insights into a model. It
should be clear at this point that our parametric considerations are not aimed
at uniform treatment of families of similar problems. Rather, we are concerned
with a formally clean treatment of parameter uncertainty.

Let us summarize the main characteristics of our approach taken here:
\begin{enumerate}
\item Our domain of computation are the real numbers in contrast to the complex
  numbers. This is the natural choice for reaction networks. It allows us to
  consequently use the information $\kk > 0$ throughout the computation. There
  is a perspective to discover further polynomial ordering inequalities in $\kk$
  with the derivation of equivalent conditions for shifted toricity, even though
  the input is purely equational.
\item We take a logic approach, using polynomial constraints, arbitrary Boolean
  combinations of these constraints, and first-order quantification. In this
  way, the logical connection between the occurring constraints is shifted from
  metamathematical reasoning to object mathematics. This ensures that human
  intuition is not mixed up with automatically derived results. The long-term
  goal is to develop robust fully automatic methods and to make corresponding
  implementations in software accessible to natural scientists. Technically, we
  employ real quantifier elimination methods, normal form computations, and
  various simplification techniques.
\item Our approach aims at the geometric shape of the real variety in contrast
  to the syntactic shape of generators of the polynomial ideal. On the one hand,
  there is a strong relation between toricity of the variety and binomiality of
  the ideal \cite{eisenbud-sturmfels-binomials}, and Gröbner bases are mature
  symbolic computation tool in this regard. The relation between toricity and
  binomiality has even been generalized to shifted toricity
  \cite{grigoriev_milman2012,GrigorievIosif:19a}. On the other hand, real
  quantifier elimination methods are an equally mature tool, and they allow to
  operate directly on the real steady state variety, which is the object of
  interest from the point of view of natural sciences. Particularly with
  parameters, order inequalities enter the stage. They should not be ignored,
  and their derivation from the ideal would not be straightforward.
\end{enumerate}

Our definitions of toricity and shifted toricity are inspired by Grigoriev and
Milman's work on \textit{binomial varieties} \cite{grigoriev_milman2012}. In
joint work with Grigoriev and others, we have systematically applied them to
both complex and real steady state varieties of reaction networks
\cite{GrigorievIosif:19a}. We have furthermore studied the connection between
complex and real shifted toricity \cite{RahkooySturm:20z}. Toric dynamical
systems have been studied by Feinberg \cite{Feinberg1972} and by Horn and
Jackson \cite{Horn1972}. Craciun et al.~\cite{craciun_toric_2009} showed that
toric dynamical systems correspond to \textit{complex balancing}
\cite{feinberg-book}. There are further definitions in the literature, where the
use of the term ``toric'' is well motivated. Gatermann et al.~considered
\textit{deformed toricity} for steady state ideals \cite{Gatermann2000}. The
exact relation between the principle of complex balancing and various
definitions of toricity has obtained considerable attention in the last years
\cite{Perez-MillanDickenstein:12a,gatermann_bernsteins_2005,muller_sign_2016}.
Complex balancing itself generalizes \textit{detailed balancing}, which has
widely been used in the context of chemical reaction networks
\cite{feinberg_stability_1987,feinberg-book,Horn1972}. Gorban et
al.~\cite{GorbanYablonski:11a,GorbanMirkes:13a} related reversibility of
chemical reactions in detailed balance to binomiality of the corresponding
varieties. Historically, the principle of detailed balancing has attracted
considerable attention in the sciences. It was used by Boltzmann in 1872 in
order to prove his H-theorem \cite{boltzmann1964lectures}, by Einstein in 1916
for his quantum theory of emission and absorption of radiation
\cite{einstein1916strahlungs}, and by Wegscheider \cite{Wegscheider1901} and
Onsager \cite{onsager1931reciprocal} in the context of \textit{chemical
  kinetics}, which led to Onsager's Nobel prize in Chemistry in 1968.
Pérez--Millán et al. \cite{Perez-MillanDickenstein:12a} consider steady state
ideals with binomial generators. They present a sufficient linear algebra
condition on the \textit{stoichiometry matrix} of a reaction network in order to
test whether the steady state ideal has binomial generators. Conradi and Kahle
proposed a corresponding heuristic algorithm. They furthermore showed that the
sufficient condition is even equivalent when the ideal is homogenous
\cite{conradi2015detecting,kahle-binomial-package-2010,Kahle2010}. Based on the
above-mentioned linear algebra condition, MESSI systems have been introduced in
\cite{millan_structure_2018}. Another linear algebra approach to binomiality has
been studied in \cite{RadulescuRahkooySturm:20}. Recently, binomiality of steady
state ideals was used to infer network structure of chemical reaction networks
out of measurement data \cite{Wang_Lin_Sontag_Sorger_2019}.

Bradford et al.~\cite{BradfordDavenport:17c,BradfordDavenport:19a} and England
et al.~\cite{EnglandErrami:17b} have worked on multistationarity of reaction
networks with parametric rate constants. Pérez--Millán et al., in their
above-mentioned work \cite{Perez-MillanDickenstein:12a}, have also discussed the
parametric case, remarkably, already in 2012. We have taken various of our
examples in the present article from \cite{Perez-MillanDickenstein:12a}, which
allows the reader to directly compare our results obtained here over the real
numbers with the existing ones over the complex numbers.

In Sect.~\ref{se:toric}, we make precise our notions of toricity and shifted
toricity. We choose a strictly formal approach leading to characterizing
first-order logic formulas over the reals. This prepares the application of real
quantifier elimination methods. In Sect.~\ref{se:qe}, we summarize basic
concepts from real quantifier elimination theory and related simplification
techniques to the extent necessary to understand our computational approach. In
Sect.~\ref{se:computations}, we present systematic computations on biological
networks taken from the literature and from established biological databases for
such models \cite{BioModels2015a}. In Sect.~\ref{se:conclusions}, we summarize
our findings and draw conclusions.

\section{Tori Are Groups, and Shifted Tori Are Cosets}\label{se:toric}
We start with some notational conventions. For a vector
$\vv = (v_1, \dots, v_n)$ equations $\vv = 0$ have to be read as
$v_1 = 0 \mathrel{\land} \dots \mathrel{\land} v_n = 0$, which is equivalent to
$\vv = (0, \dots, 0)$. Inequalities $\vv \neq 0 $ have to be read as
$v_1 \neq 0 \mathrel{\land} \dots \mathrel{\land} v_n \neq 0$, which is \emph{not} equivalent to
$\vv \neq (0, \dots, 0)$. Similarly, inequalities $\vv > 0$ serve as shorthand
notations for $v_1 > 0 \mathrel{\land} \dots \mathrel{\land} v_n > 0$. Other ordering
relations will not occur with vectors. All arithmetic on vectors is
component-wise. Logic formulas as above are mathematical objects that can
contain equations. For better readability we use ``$\dotequal$'' to express
equality between formulas.

Consider polynomials $\ff \in \ZZ[\kk, \xx]^m$ with parameters
$\kk = (k_1, \dots, k_s)$ and variables $\xx = (x_1, \dots, x_n)$. For fixed choices
$\kk^* \in \Rp{s}$ of $\kk$, the corresponding real variety of $\ff$ is given by
\begin{equation}
  \Vks(\ff) = \{\,\xx^* \in \RR^n \mid \ff(\kk^*,\xx^*) = 0\,\}.
\end{equation}
We consider the multiplicative group $\RR^* = \RR \setminus \{0\}$, note that the direct
product $\Rsn$ establishes again a group, and define
\begin{equation}
  \Vks(\ff)^* = \Vks(\ff) \cap \Rsn \subseteq \Rsn.
\end{equation}
This set $\Vks(\ff)^*$ is a \emph{torus} if it forms an irreducible subgroup of
$\Rsn$. For this purpose, we allow ourselves to call $\Vks(\ff)^*$ irreducible
if $\Vks(\ff)$ is irreducible, equivalently, if
$\langle \ff(\kk^*, \xx) \rangle$ is a prime ideal over $\RR$. More generally,
$\Vks(\ff)^*$ is a \emph{shifted torus} if it forms an irreducible coset of
$\Rsn$ \cite{grigoriev_milman2012,GrigorievIosif:19a}.

In this article, we focus on the discovery of coset and group structures. This
is only a very mild limitation, as a closer look at the geometric relevance of
the irreducibility requirement shows: If we discover a coset but irreducibility
does not hold, then we are, from a strictly geometrical point of view, faced
with finitely many shifted tori instead of a single one. If we disprove the
coset property and irreducibility does not hold, then some but not all of the
irreducible components might be shifted tori, and they could be discovered via
decomposition of the variety. The same holds for groups vs.~tori.

It should be noted that the primality of $\langle \ff(\kk^*, \xx) \rangle$ over
$\RR$ in contrast to $\QQ$ is a computationally delicate problem already in the
non-parametric case. Starting with integer coefficients, prime decomposition
would require the construction of suitable real extension fields during
computation. Our parametric setting would require in addition the introduction
of suitable finite case distinctions on the vanishing of coefficient polynomials
in $\kk$.

The definition typically used for a set $C \subseteq \Rsn$ to form a coset of
$\Rsn$ goes as follows: There exists $\gggg \in \Rsn$ such that $\gggg^{-1} C$
forms a subgroup of $\Rsn$. We are going to use a slightly different but
equivalent characterization: $\gggg^{-1} C$ forms a subgroup of $\Rsn$ for all
$\gggg \in C$. A proof for the equivalence can be found in \cite[Proposition
21]{GrigorievIosif:19a}. We now present four first-order logic formulas $\PHI1$,
\dots,~$\PHI4$. They state, uniformly in $\kk$, certain properties that can be
combined to express that $\Vk(\ff)^*$ forms a coset or a group:
\begin{enumerate}
\item \textit{Non-emptiness}\\
  There exists $\xx \in \Rsn$ such that $\xx \in\Vk(\ff)$:
  \begin{equation}\label{eq:phi1}
    \PHI1 \dotequal \exists \xx (\xx \neq 0 \land \ff = 0).
  \end{equation}

\item \textit{Shifted completeness under inverses}\\
  For all $\gggg$, $\xx \in \Rsn$, if $\gggg$, $\gggg\xx \in \Vk(\ff)$, then
  $\gggg\xx^{-1} \in \Vk(\ff)$:
  \begin{multline}\label{eq:phi2}
    \PHI2 \dotequal
    \forall \gggg \forall \xx (\gggg \neq 0 \land \xx \neq 0 \land \ff[\xx \gets \gggg] = 0 \land
    \ff[\xx \gets \gggg\cdot\xx] = 0  {}\\ \longrightarrow \ff[\xx \gets \gggg\cdot\xx^{-1}] = 0).
  \end{multline}
  Here $[\xx \gets \tttt]$ denotes substitution of terms $\tttt$ for variables
  $\xx$. In the equation $\ff[\xx \gets \gggg\cdot\xx^{-1}] = 0$ we tacitly drop the
  principal denominator of the left hand side to obtain a polynomial. This is
  admissible due to the premise that $\xx \neq 0$.\smallskip

\item \textit{Shifted completeness under multiplication}\\
  For all $\gggg$, $\xx$, $\yy \in \Rsn$, if $\gggg$, $\gggg\xx$,
  $\gggg\yy \in \Vk(\ff)$, then $\gggg\xx\yy \in \Vk(\ff)$:
  \begin{multline}\label{eq:phi3}
    \PHI3 \dotequal {}
    \forall \gggg \forall \xx \forall \yy (\gggg \neq 0 \land \xx \neq 0 \land \yy \neq 0 \land
    \ff[\xx \gets \gggg] = 0 \land {}\\
    \ff[\xx \gets \gggg\cdot\xx] = 0 \land \ff[\xx \gets \gggg\cdot\yy] = 0
    \longrightarrow \ff[\xx \gets \gggg\cdot\xx\cdot\yy] = 0).
  \end{multline}

\item \textit{Neutral element}\\
  $(1, \dots, 1) \in \Vk(\ff)$:
  \begin{equation}\label{eq:phi4}
    \PHI4 \dotequal \ff[\xx \gets (1, \dots, 1)] = 0.
  \end{equation}
\end{enumerate}

In these terms we can define formulas $\sigma$ and $\tau$, which state the $\Vk(\ff)^*$
is a coset or group, respectively:
\begin{equation}\label{eq:sigmatau}
  \sigma \dotequal \PHI1 \land \PHI2 \land \PHI3,\quad
  \tau \dotequal \PHI2 \land \PHI3 \land \PHI4.
\end{equation}

For the non-parametric case, these formulas have been derived and discussed in
\cite[Sect.~3.2]{GrigorievIosif:19a}. In the absence of parameters they were
logic sentences, which are either true or false over the real numbers. Real
decision produces were used to automatically derive either ``$\true$'' or
``$\false$.'' In our parametric setting here, they contain $\kk$ as free
variables and thus establish exact formal conditions in $\kk$, which become
either ``$\true$'' or ``$\false$'' after making choices of real values for
$\kk$.

\section{Real Quantifier Elimination and Simplification}\label{se:qe}
In the presence of parameters, the natural generalization of a decision
procedure is an effective \emph{quantifier elimination} procedure for the real
numbers \cite{DolzmannSturm:98a,Sturm:17a,Sturm_ISSAC2018}. In fact, most real
decision procedures are actually quantifier elimination procedures themselves,
which apply quantifier elimination to their parameter-free input and
subsequently evaluate the variable-free quantifier elimination result to either
``$\true$'' or ``$\false$.'' Plenty of approaches have been proposed for real
quantifier elimination,
e.g.~\cite{Tarski:48a,CollinsHong:91,Grigoriev:88a,LoosWeispfenning:93a,BasuPollack:96a,Weispfenning:97b,Weispfenning:98a,Kosta:16a},
but only few of them have led to publicly available implementations with a
long-term support strategy
\cite{DolzmannSturm:97a,Brown:03a,DBLP:journals/jsc/Strzebonski06,DBLP:journals/jsc/ChenM16,Tonks:20f}.

Given a first-order formula $\varphi$ built from polynomial constraints with integer
coefficients, quantifier elimination computes a formula $\varphi'$ that is equivalent
to $\varphi$ over the reals, formally
$\RR \models \varphi \longleftrightarrow \varphi'$, but does not contain any quantifiers. We allow ourselves to call
$\varphi'$ \emph{the result} of the quantifier elimination, although it is not
uniquely determined by $\varphi$.

The following example, which is discussed in more detail in
\cite[Sect.~2.1]{Sturm:17a}, gives a first idea: On input of
\begin{equation}
  \varphi \dotequal \forall x_1 \exists x_2(x_1^2+x_1x_2+k_2 > 0 \land x_1+k_1x_2^2+k_2 \leq 0),
\end{equation}
quantifier elimination computes the result
$\varphi' \dotequal k_1 < 0 \land k_2 > 0$, which provides a necessary and sufficient
condition in $\kk$ for $\varphi$ to hold. Another application of quantifier
elimination has been used already in the introduction of this article: Consider
$\ff = (f_1, f_3, f_4)$ with $f_1$, $f_3$, $f_4$ as in \eqref{eq:pho1kin}. Then
compute $\PHI2$, \dots,~$\PHI4$ as in \eqref{eq:phi2}--\eqref{eq:phi4} and
$\tau$ as in \eqref{eq:sigmatau}. On input of $\tau$, quantifier elimination delivers
the result
$\tau' \dotequal \kon{} - \koff{} = \lon{} - \loff{} = \kcat{} = \lcat{}$. This is
a necessary and sufficient condition in $\kk$ for $\Vk(\ff)^*$ to form a group,
which has already been presented in (\ref{it:npho1ii}) on
p.\pageref{it:npho1ii}.

For an existential formula like $\varphi_1$ in \eqref{eq:phi1}, quantifier elimination
computes a result $\varphi_1'$ that provides necessary and sufficient conditions in
$\kk$ for the existence of choices for $\xx$ that satisfy the constraints in
$\varphi_1$. By definition, quantifier elimination does not derive any information on
possible choices of $\xx$. In other words, quantifier elimination talks about
solvability, not about solutions. However, quantifier elimination via virtual
substitution
\cite{LoosWeispfenning:93a,Weispfenning:97b,Kosta:16a,Sturm_ISSAC2018}, which we
use here primarily, can optionally provide sample solutions for $\xx$. This is
known as \emph{extended quantifier elimination} \cite{KostaSturm:16a}. We have
used extended quantifier elimination to compute the uniform steady state $\xx^*$
in \eqref{eq:pho1ss} in the introduction, besides the actual quantifier
elimination result ``$\true$.''

Successful practical application of quantifier elimination by virtual
substitution goes hand in hand with strong and efficient automatic
simplification of intermediate and finite results. We use essentially a
collection of techniques specified in \cite[Sect.~5.2]{DolzmannSturm:97c} as the
``standard simplifier.'' In particular, we exploit the concept of an
\emph{external theory} introduced in \cite{DolzmannSturm:97c} with $\kk > 0$ as
our theory. This means that all simplifications are performed modulo the
assumption $\kk > 0$ without explicitly adding this information to the input
formula $\varphi$. As a consequence, the quantifier elimination result
$\varphi'$ is equivalent only modulo $\kk > 0$, formally
$\RR \models \kk > 0 \longrightarrow (\varphi \longleftrightarrow \varphi')$.\footnote{Alternatively, one could temporarily
  introduce constants $\kk$ and state equivalence in an extended theory of real
  closed fields:
  $\operatorname{Th}(\RR) \cup \{\kk>0\} \models \varphi \longleftrightarrow \varphi'$. This point of view is common in
  algebraic model theory and has been taken in \cite{DolzmannSturm:97c}.}

Note that, in contrast to $\kk > 0$ for the rate constants, we never require
$\xx > 0$ for the species concentrations although chemical reaction network
theory assumes both to be positive. The reason is that the concepts of toricity
used here have been defined in terms of varieties and multiplicative groups
without any reference to order. It might be interesting to review these concepts
with respect to the particular situation encountered here. However, this is
beyond the scope of this article and should be settled in a non-parametric
context first.

We convert our final results to disjunctive normal form
\cite[Sect.~7]{DolzmannSturm:97c} and apply simplification methods based on
Gröbner bases \cite[Sect.~4.3]{DolzmannSturm:97c}. A disjunctive normal form is
a finite case distinction over systems of constraints. It has been our
experience that users prefer such a presentation of the computed information in
comparison to arbitrary boolean combinations, even at the price of larger
output. In general, this normal form computation can get quite expensive in time
and space, because quantifier elimination by virtual substitution on universal
formulas like $\PHI2$, \dots,~$\PHI4$ in \eqref{eq:phi2}--\eqref{eq:phi4} tends to
produce conjunctions of disjunctions rather than vice versa. Luckily, our
results are rather small.

Having said this, we have devised \emph{quantifier elimination-based
  simplification} as another heuristic simplification step for our results
$\psi$ here. It checks via quantifier elimination for every single constraint
$\gamma$ in $\psi$ whether
\begin{equation}\label{eq:qbs}
  \RR \models \forall\kk(\kk > 0 \longrightarrow \gamma) \longleftrightarrow \true\quad\text{or}\quad
  \RR \models \exists\kk(\kk > 0 \land \gamma) \longleftrightarrow \false.
\end{equation}
When such constraints $\gamma$ are found, they are replaced in $\psi$ with the
respective truth value, and then the standard simplifier in applied to $\psi$ once
more. Quantifier elimination-based simplification preserves disjunctive normal
forms.

As an example consider
$\kk = (k_{12}, k_{13}, k_{21}, k_{23}, k_{31}, k_{32})^T$ and
$\psi \dotequal \gamma_1 \lor \gamma_2$, where
\begin{align*}
  \gamma_1 & \dotequal k_{31} - k_{32} = 0\\
  \gamma_2 & \dotequal 16 k_{12} k_{21} + 8 k_{12} k_{23} + 8 k_{13} k_{21} + 4
        k_{13} k_{23} +   k_{31}^{2} - 2 k_{31} k_{32} + k_{32}^{2} \leq
        0.\numberthis\label{eq:gamma2} 
\end{align*}
If one recognizes that
$k_{31}^{2} - 2 k_{31} k_{32} + k_{32}^{2} = (k_{31} - k_{32})^{2}$ and
furthermore takes into consideration that $\kk > 0$, it becomes clear that
$\gamma_2$ is not satisfiable. The argument can be seen as a generalization of
sum-of-squares decomposition, which is not supported within our simplification
framework \cite{DolzmannSturm:97c}. Quantifier elimination-based simplification
recognizes that the condition on the right hand side of \eqref{eq:qbs} holds for
$\gamma_2$. It replaces $\gamma_2$ with ``$\false$'' in $\psi$, which yields
$\gamma_1 \lor \false$. Finally, the standard simplifier is applied, and
$\gamma_1$ is returned.

\section{Computational Experiments}\label{se:computations}
All our computations have been conducted on an AMD EPYC 7702 64-Core Processor.
On the software side, we have used SVN revision 5797 of the computer algebra
system Reduce with its computer logic package Redlog
\cite{10.1145/2402536.2402544,Hearn:05a,DolzmannSturm:97a}. Reduce is open
source and freely available on
SourceForge.\footnote{\url{https://sourceforge.net/projects/reduce-algebra/}} On
these grounds, we have implemented systematic Reduce scripts, which essentially
give algorithms and could be turned into functions as a next step. In few
places, global Redlog options have been adjusted manually in order to optimize
the efficiency of quantifier elimination for a particular example. The scripts
and the log files of the computations are available as supplementary material
with this article.

\subsection{An Artificial Triangle Network}\label{se:triangle}
We start with an artificial network introduced by Pérez--Millán et
al.~\cite[p.1033, Ex.~2.3]{Perez-MillanDickenstein:12a}:
\begin{equation}\label{eq:trianglenet}
  \ce{2A <=>[$k_{12}$][$k_{21}$] 2B <=>[$k_{23}$][$k_{32}$] A + B
    <=>[$k_{31}$][$k_{13}$] 2A}.
\end{equation}
There are reaction rates
$\kk = (k_{12}, k_{13}, k_{21}, k_{23}, k_{31}, k_{32})^T$ and species
concentrations $\xx = (x_1, x_2)^T$ for abstract species $\ce{A}$ and $\ce{B}$,
respectively. Its kinetics is described by an ODE $\dot{\xx} = \ff$ with a
polynomial vector field $\ff = (f_1, f_2)^T$ as follows:
\begin{equation}
  f_1 = f_2 = (-2k_{12}-k_{13}) x_{1}^{2}
  +(2 k_{21}+k_{23}) x_{2}^{2}
  +(k_{31} -k_{32}) x_{1} x_{2}.
\end{equation}

We form $\PHI1$ according to \eqref{eq:phi1}, and extended quantifier
elimination yields ${\PHIP1 \dotequal \true}$ along with a uniform witness
\begin{equation}
  \textstyle
  \xx^* = \left(
    1,
    -\frac{
      \sqrt{
        16 k_{12} k_{21} + 8 k_{12} k_{23} + 8 k_{13} k_{21} + 4 k_{13} k_{23} +
        k_{31}^{2} - 2 k_{31} k_{32} + k_{32}^{2}} - k_{31} + k_{32
      }
    }{
      4 k_{21} + 2 k_{23}
    }
  \right)^T.
\end{equation}
Notice that $\kk > 0$ ensures that the denominator cannot vanish.

Next, we consider $\PHI2$ and obtain $\PHIP2 \dotequal k_{31} - k_{32} = 0$ with
the help of quantifier elimination-based simplification. In fact, this is the
example for quantifier elimination-based simplification discussed in the
previous section. From $\PHI3$ we also obtain
$\PHIP3 \dotequal k_{31} - k_{32} = 0$.

Hence, $\Vk(\ff)^*$ forms a coset of $\Rs2$ if and only if $\RR \models \sigma'$, where
\begin{equation}
  \sigma' = k_{31} - k_{32} = 0.
\end{equation}
The same condition has been derived with a different method in
\cite{Perez-MillanDickenstein:12a}. For $\Vk(\ff)^*$ to form even a subgroup of
$\Rs2$ we must add to $\sigma'$ the condition
$\PHI4 \dotequal \ff[\xx \gets (1,1)] = 0$. This yields
\begin{equation}
  \tau' \dotequal k_{31} - k_{32} = 0 \land 2 k_{12} + k_{13} - 2 k_{21} - k_{23} = 0.
\end{equation}

The overall CPU time for the computations in this section was 0.867~s. Details
on input problem sizes can be found in Tab.~\ref{tab:other}.
\begin{table}[t]
  \centering
  \caption{Problem sizes and computation times for
    Sect.~\protect\ref{se:triangle}--Sect.~\protect\ref{se:tgfbeta}\label{tab:other}}
  \setlength{\tabcolsep}{1em}
  \begin{tabular}{llrrrrrrr}
    \hline
    Sect. & network & $|\kk|$ & $|\xx|$ & $|\ff|$ & \multicolumn{3}{c}{\# quantifiers} & time\\
          &         &         &         &          & $\PHI1$ & $\PHI2$ & $\PHI3$ & \\
    \hline
    \ref{se:triangle} & Triangle & 6 & 2 & 2 & 2 & 4 & 6 & 0.845 s\\
    \ref{se:shinar} & EnvZ-OmpR & 14 & 9 & 9 & 9 & 18 & 27 & 2.172 s\\
    \ref{se:tgfbeta} & TGF-$\beta$ & 8 & 6 & 6 & 6 & 12 & 18 & 26.477 s\\
    \hline
  \end{tabular}
\end{table}

\subsection{Escherichia Coli Osmoregulation System}\label{se:shinar}
Our next example is a model of the escherichia coli osmoregulation system
(EnvZ-OmpR). It has been introduced by Shinar and Feinberg \cite[(S60) in the
supporting online material]{ShinarFeinberg:10z} and also discussed in
\cite[p.1043, Example 3.15]{Perez-MillanDickenstein:12a}:
\begin{align*}
  &\ce{XD <=>[$k_{12}$][$k_{21}$] X <=>[$k_{23}$][$k_{32}$] XT ->[$k_{34}$] X_P}
  &&\ce{XT + Y_P <=>[$k_{89}$][$k_{98}$] XTY_P ->[$k_{9,10}$] XT + Y}\\
  &\ce{X_P + Y <=>[$k_{56}$][$k_{65}$] X_PY ->[$k_{67}$] X + Y_P}\quad
  &&\ce{XD + Y_P <=>[$k_{11,12}$][$k_{12,11}$] XDY_P ->[$k_{12,13}$] XD + Y}.\numberthis
\end{align*}
There are 14 reaction rates $\kk$ and species concentrations
$\xx = (x_1, \dots, x_9)^T$ for $\ce{XD}$, $\ce{X}$, $\ce{XT}$, $\ce{X_P}$,
$\ce{Y}$, $\ce{X_PY}$, $\ce{Y_P}$, $\ce{XTY_P}$, $\ce{XDY_P}$, respectively. Its
kinetics is described by an ODE $\dot{\xx} = \ff$ with a polynomial vector field
$\ff = (f_1, \dots, f_9)^T$ as follows:
\begin{align*}
  f_1 & =  -k_{12} x_1 + k_{21}x_2 - k_{11,12}x_1x_7 + (k_{12,11}+k_{12,13})x_9\\
  f_2 & = k_{12} x_{1} + (-k_{21}-k_{23}) x_{2} + k_{32} x_{3} + k_{67} x_{6}\\
  f_3 & = k_{23} x_{2} + (-k_{32}-k_{34}) x_{3} - k_{89} x_{3} x_{7} +
        (k_{98}+k_{9,10}) x_{8}\\
  f_4 & = k_{34} x_{3} - k_{56} x_{4} x_{5} + k_{65} x_{6}\\
  f_5 & = -k_{56} x_{4} x_{5} + k_{65} x_{6} + k_{9,10} x_{8} + k_{12,13} x_{9}\\
  f_6 & = k_{56} x_{4} x_{5} + (-k_{65}-k_{67}) x_{6}\\
  f_7 & = k_{67} x_{6} - k_{89} x_{3} x_{7} + k_{98} x_{8} - k_{11,12} x_{1}
        x_{7} + k_{12,11} x_{9}\\
  f_8 & = k_{89} x_{3} x_{7} + (-k_{98}-k_{9,10}) x_{8}\\
  f_9 & = k_{11,12} x_{1} x_{7} + (-k_{12,11}-k_{12,13}) x_{9}.\numberthis
\end{align*}

We compute
$\PHIP1 \dotequal \PHIP2 \dotequal \PHIP3 \dotequal \sigma \dotequal \true$, which
means that $\Vk(\ff)^*$ forms a coset for all admissible choices of reaction
rates $\kk$. Again, extended quantifier elimination delivers, in addition to
$\PHIP1$, a uniform parametric witness $\xx^*$ for the non-emptiness of
$\Vk(\ff)^*$. We obtain the following equivalent condition in $\kk$ for
$\Vk(\ff)^*$ to form even a group:
\begin{align*}
  \PHIP4 \ \dotequal \ \tau \ \dotequal \ {}
  & k_{89} - k_{9,10} - k_{98} = 0 \land k_{12,13} - k_{67} + k_{89} - k_{98} = 0 \land{}\\
  & k_{12,13} - k_{56} + k_{65} + k_{89} - k_{98} = 0 \land k_{12,13} - k_{34} +
    k_{89} - k_{98} = 0 \land{}\\
  & k_{12,13} - k_{23} + k_{32} + k_{89} - k_{98} = 0 \land k_{12} - k_{21} = 0 \land{}\\
  & k_{11,12} - k_{12,11} - k_{12,13} = 0.\numberthis
\end{align*}

The overall CPU time for the computations in this section was 0.651~s. Details
on input problem sizes can be found in Tab.~\ref{tab:other}.
 
\subsection{TGF-$\beta$ Pathway}\label{se:tgfbeta}
The TGF-$\beta$ signaling pathway plays a central role in tissue homeostasis and
morphogenesis, as well as in numerous diseases such as fibrosis and cancer
\cite{VilarJansen:06}. It is featured as model no.~101 in the BioModels
repository
\cite{BioModels2015a}.\footnote{\url{https://www.ebi.ac.uk/biomodels/BIOMD0000000101}}
We consider here a variant, which ignores a discrete event changing \ce{ligand}
concentration at time $t=2500$. A non-parametric instance of this variant has
been studied in \cite{GrigorievIosif:19a} with respect to toricity and in
\cite{KruffLueders:21} with respect to multiple time scale reduction.
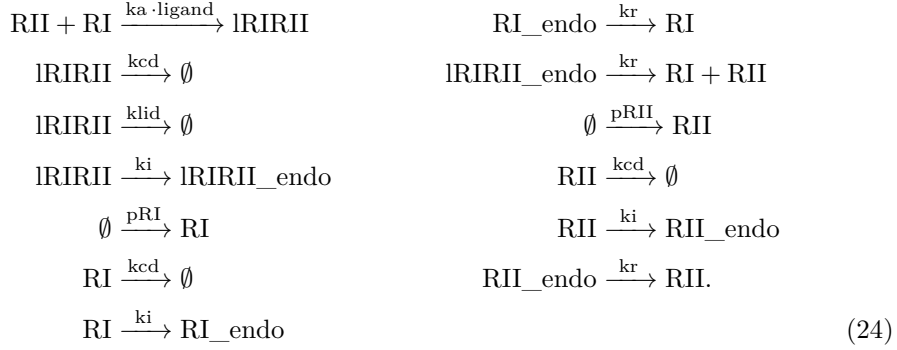
\begin{align*}
  \ce{RII + RI &->[ka \cdot ligand] lRIRII}
  & \ce{RI\_endo &->[kr] RI}\\
  \ce{lRIRII &->[kcd] \emptyset}
  & \ce{lRIRII\_endo &->[kr] RI + RII}\\
  \ce{lRIRII &->[klid] \emptyset}
  & \ce{\emptyset &->[pRII] RII}\\
  \ce{lRIRII &->[ki] lRIRII\_endo}
  & \ce{RII &->[kcd] \emptyset}\\
  \ce{\emptyset &->[pRI] RI}
  & \ce{RII &->[ki] RII\_endo}\\
  \ce{RI &->[kcd] \emptyset}
  & \ce{RII\_endo &->[kr] RII}.\\
  \ce{RI &->[ki] RI\_endo}\numberthis\label{eq:tgfbetanet}
\end{align*}
There are 8 parameters $\kk$ and species concentrations
$\xx = (x_1, \dots, x_6)^T$ corresponding to \ce{RI}, \ce{RII}, \ce{lRIRII},
\ce{lRIRII\_endo}, \ce{RI\_endo}, \ce{RII\_endo}, respectively. The dynamics of
the network is described by an ODE $\dot{\xx} = \ff$ with a polynomial vector
field $\ff = (f_1, \dots, f_6)^T$ as follows:
\begin{align*}
  f_1 &= - \ce{ka} \cdot \ce{ligand} \cdot x_{1} x_{2} - \ce{kcd} \cdot x_{1} - \ce{ki}
        \cdot x_{1} + \ce{kr} \cdot x_{4} + \ce{kr} \cdot x_{5} + \ce{pri}\\
  f_2 &= - \ce{ka} \cdot \ce{ligand} \cdot x_{1} x_{2} - \ce{kcd} \cdot x_{2} - \ce{ki}
        x_{2} + \ce{kr} \cdot x_{4} + \ce{kr} \cdot x_{6} + \ce{prii}\\
  f_3 &= \ce{ka} \cdot \ce{ligand} \cdot x_{1} x_{2} - \ce{kcd} \cdot x_{3} - \ce{ki} \cdot
        x_{3} - \ce{klid} \cdot x_{3}\\
  f_4 &= \ce{ki} \cdot x_{3} - \ce{kr} \cdot x_{4}\\
  f_5 &= \ce{ki} \cdot x_{1} - \ce{kr} \cdot x_{5}\\
  f_6 &= \ce{ki} \cdot x_{2} - \ce{kr} \cdot x_{6}.\numberthis\label{eq:tgfbetakin}
\end{align*}

For fixed choices $\kk^*$ of parameters as specified in the BioModels repository
we had shown in \cite{GrigorievIosif:19a} that $\Vks(\ff)^*$ is not a coset. Our
parametric approach here allows to investigate to what extent this negative
result depends on the specific choices $\kk^*$. We compute
$\PHIP1 \dotequal \true$ along with a witness for $\Vk(\ff)^* \neq \emptyset$ for all
admissible choices of $\kk$. Next, we obtain
$\PHIP2 \dotequal \PHIP3 \dotequal \false$, i.e., shifted completeness under
inverses and multiplication fails for all admissible choices of $\kk$. It
follows that $\sigma \dotequal \tau \dotequal \false$, i.e., $\Vk(\ff)^*$ is generally
not a coset and not a group.

The synthesis and degradation reactions\footnote{i.e., the ones with
  ``$\emptyset$'' on their left hand side or right hand side, respectively} in
\eqref{eq:tgfbetanet} cause absolute summands in $f_1$ and $f_2$ in the dynamics
\eqref{eq:tgfbetakin}. Although there is a connection between cosets and the
existence of binomial generators of the ideal, those summands are not an
immediate reason to exclude cosets. Consider, e.g., the abstract example
$\gggg = (-x_1 - x_2 + k_1, x_2 + x_3+ k_2)$, where $\Vk(\gggg)^*$ is a coset
for all admissible choices of $k_1$, $k_2$. On the other hand, we have mentioned
in the introduction that toricity is related to complex balance. TGF-$\beta$ cannot
not have complex balance, because there is a nonzero flux through the system:
receptors are produced, they cycle, and are degraded. One cannot transfer
information without dissipation. This observation generally applies to signaling
models.

The overall CPU time for the computations in this section was 26.477~s. Details
on input problem sizes can be found in Tab.~\ref{tab:other}.

\subsection{n-Site Phosphorylation-Dephosphorylation Cycle}\label{se:npho}
The $n$-site phosphorylation network in the form discussed here has been taken
from Wang and Sontag~\cite{WangSontag:08p}. Pérez--Millán et al.~have discussed
$n$-site phosphorylation for generic $n$
\cite[Sect.~4.1]{Perez-MillanDickenstein:12a}; the cases $n=1$ and $n=2$ are
discussed explicitly as Ex.~2.1 and Ex.~3.13, respectively. We have used the
case $n=1$ in the introduction.

For a fixed positive integer $n$, the $n$-site phosphorylation reaction network
is given by
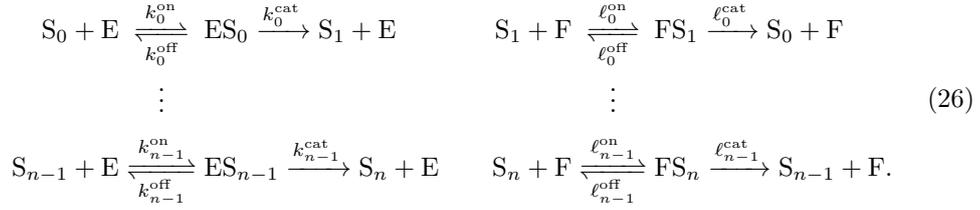
\begin{equation}
  \renewcommand{\arraystretch}{2}
  \setlength{\arraycolsep}{0pt}
  \begin{array}{rcl@{\quad\quad}rcl}
    \ce{S_0 + E & <=>[$\kon0$][$\koff0$] & ES_0 ->[$\kcat0$] S_1 + E} &
    \ce{S_1 + F & <=>[$\lon0$][$\loff0$] & FS_1 ->[$\lcat0$] S_0 + F}\\
    & \vdots & &  & \vdots &\\
    \ce{S_{n-1} + E & <=>[$\kon{n-1}$][$\koff{n-1}$] & ES_{n-1}
    ->[$\kcat{n-1}$] S_n + E} &
    \ce{S_n + F & <=>[$\lon{n-1}$][$\loff{n-1}$] & FS_n
    ->[$\lcat{n-1}$] S_{n-1} + F}.
  \end{array}
\end{equation}
Its dynamics is described by the following ODE with $6n$ parameters
$\kk_n = (\kon0, \dots, \lcat{n-1})$ and $3n+3$ variables
\begin{equation}
  \xx_n = (s_0, \dots, s_n, c_0, \dots, c_{n-1}, d_1, \dots, d_n, e, f)
\end{equation}
for concentrations of species $\ce{S_0}$, \dots,~$\ce{S_n}$, $\ce{ES_0}$,
\dots,~$\ce{ES_{n-1}}$, $\ce{FS_1}$, \dots,~$\ce{FS_n}$, $\ce{E}$, $\ce{F}$,
respectively:
\begin{align*}
  \dot{s_0} & = -\kon0 s_0 e + \koff0 c_0 + \lcat0 d_1\\
  \dot{s_i} & = -\kon{i} s_i e + \koff{i} c_i + \kcat{i-1} c_{i-1} -
                \lon{i-1} s_i f + \loff{i-1} d_i + \lcat{i} d_{i+1}\\
  \dot{c_j} & = \kon{j} s_j e - (\koff{j} + \kcat{j}) c_j\\
  \dot{d_k} & =\lon{k-1} s_k f - (\loff{k-1} + \lcat{k-1}) d_k,\\
            & \quad\quad\quad\quad\quad\quad\quad i = 1, \dots, n-1,\quad j=0, \dots, n-1,\quad k=1, \dots, n.
              \numberthis\label{eq:npho}
\end{align*}
Let $\ff_n = (f_1, \dots, f_{3n-1})$ denote the vector field of \eqref{eq:npho}. We
may ignore here the equations for $\dot{s_n}$, $\dot{e}$, and $\dot{f}$, whose
right hand sides are linear combinations of $\ff_n$.

For $n \in \{1, \dots, 5\}$ we obtain the following computational results:
\begin{enumerate}[(i)]
\item $\Vk(\ff)^* \neq \emptyset$ for all admissible choices of $\kk$; we also obtain a
  uniform witness in terms of $\kk$;
\item $\Vk(\ff)^*$ forms a coset for all admissible choices of $\kk$;
\item $\Vk(\ff)^*$ forms a group if and only if
  \begin{equation}\label{eq:nphogroup}
    \bigwedge_{i=0}^{n-1}\kon{i} - \koff{i} = \lon{i} - \loff{i} = \kcat{i} = \lcat{i}.
  \end{equation}
\end{enumerate}

Wang and Sontag, in their article \cite{WangSontag:08p}, were interested in
quantitative information on the numbers of steady states of the dynamics
\eqref{eq:npho}. Our results here provide qualitative information on the
structure of the set of steady states. We could automatically deduce that there
is always at least one steady state, for which we find a uniform witness in
$\kk$. In fact, extended quantifier elimination could even enumerate steady
states, because one can exclude in the input formula the ones already found, and
rerun. More important, we know that the set $S \subseteq \Rsn$ of all steady states
forms a coset. That is, for all choices of $\kk$ and all $\gggg \in S$, the set
$G = \gggg^{-1} S$ is complete under component-wise multiplication and inverses.
The set $S$ itself has this completeness property only for choices of parameters
satisfying the equations \eqref{eq:nphogroup} exactly, which one cannot expect
from a practical point of view.

As one possible application of our results, assume that experiments have
delivered three steady states $\xx_1$, \dots,~$\xx_3$. Then, e.g., the following are
steady states, too: \begin{equation}
  \xx_1(\xx_1^{-1}\xx_2\cdot\xx_1^{-1}\xx_3)=\xx_1^{-1}\xx_2\xx_3,\quad
  \xx_1(\xx_1^{-1}\xx_2)^{-1} = \xx_1^2\xx_2.
\end{equation}
Here we use multiplication with $\xx_1^{-1}$ for switching from $S$ to $G$,
exploit there completeness under multiplication and inverses, respectively, and
finally use multiplication with $\xx_1$ for switching back to $S$.

The computation times are collected in Tab.~\ref{tab:npho}.
\begin{table}[t]
  \centering
  \caption{Problem sizes and computation times for n-site phosphorylation in
    Sect.\protect\ref{se:npho}\label{tab:npho}}
  \setlength{\tabcolsep}{0.75em}
  \begin{tabular}{rrrrrrrr}
    \hline
    $n$ & $|\kk|$ & $|\xx|$ & $|\ff|$ & \multicolumn{3}{c}{\# quantifiers} & time\\
        &         &         &          & $\PHI1$ & $\PHI2$ & $\PHI3$ & \\
    \hline
    $1$ & 6 & 6 & 2 & 6 & 12 & 18 & 0.500 s\\
    $2$ & 12 & 9 & 5 & 9 & 18 & 27 & 1.131 s\\
    $3$ & 18 & 12 & 8 & 12 & 24 & 36 & 5.911 s\\
    $4$ & 24 & 15 & 11 & 15 & 30 & 45 & 33.790 s\\
    $5$ & 30 & 18 & 14 & 18 & 36 & 54 & 3963.204 s\\
    $\geq6$ & $6n$ & $3(n+1)$ & $3n-1$ & $3(n+1)$ & $6(n+1)$ & $9(n+1)$ & $>$ 6 h\\
    \hline
  \end{tabular}
\end{table}
The formula $\PHI3$ for $n=5$ is the formally largest quantifier elimination
problem considered in this article. We have eliminated here 54 real quantifiers
in an 84-dimensional space, which took 1 h 6 min. For $n \geq 6$, the computations
did not finish within 6 hours.

\subsection{Excitatory Post-Synaptic Potential Acetylcholine
  Event}\label{se:biomd01}
The excitatory post-synaptic potential acetylcholine event model (EPSP-ACh) has
been introduced by Edelstein et al.~\cite{EdelsteinSchaad:96r}. It also appears
as model no.~1 in the BioModels repository
{\cite{BioModels2015a}}:\footnote{\url{https://www.ebi.ac.uk/biomodels/BIOMD0000000001}}
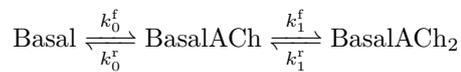
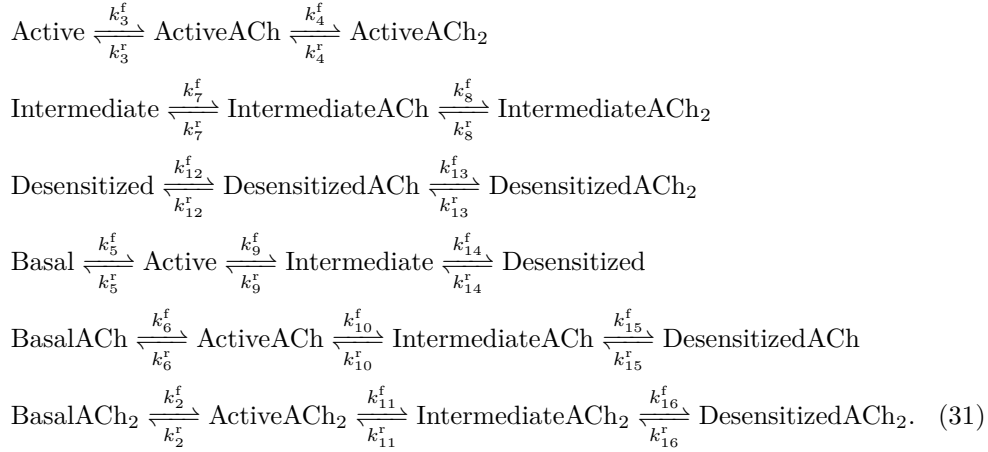
\begin{align*}
  &\ce{Basal <=>[$\kf0$][$\kr0$] BasalACh <=>[$\kf1$][$\kr1$] BasalACh2}\\
  &\ce{Active <=>[$\kf3$][$\kr3$] ActiveACh <=>[$\kf4$][$\kr4$] ActiveACh2}\\
  &\ce{Intermediate <=>[$\kf7$][$\kr7$] IntermediateACh <=>[$\kf8$][$\kr8$] IntermediateACh2}\\
  &\ce{Desensitized <=>[$\kf{12}$][$\kr{12}$] DesensitizedACh
    <=>[$\kf{13}$][$\kr{13}$] DesensitizedACh2}\\
  &\ce{Basal <=>[$\kf5$][$\kr5$] Active <=>[$\kf9$][$\kr9$] Intermediate
    <=>[$\kf{14}$][$\kr{14}$] Desensitized}\\
  &\ce{BasalACh <=>[$\kf6$][$\kr6$] ActiveACh <=>[$\kf{10}$][$\kr{10}$]
    IntermediateACh <=>[$\kf{15}$][$\kr{15}$] DesensitizedACh}\\
  &\ce{BasalACh2 <=>[$\kf2$][$\kr2$] ActiveACh2 <=>[$\kf{11}$][$\kr{11}$]
    IntermediateACh2 <=>[$\kf{16}$][$\kr{16}$] DesensitizedACh2}.\numberthis
\end{align*}
There are 34 reaction rates $\kk$ and species concentrations
$\xx = (x_1, \dots, x_{12})^T$ for $\ce{BasalACh2}$, $\ce{IntermediateACh}$,
$\ce{ActiveACh}$, $\ce{Active}$, $\ce{BasalACh}$, $\ce{Basal}$,
$\ce{De\-sen\-si\-tizedACh2}$, $\ce{Desensitized}$, $\ce{IntermediateACh2}$,
$\ce{DesensitizedACh}$, $\ce{Intermediate}$, $\ce{ActiveACh2}$, respectively.
The kinetics is described by an ODE $\dot{\xx} = \ff$ with a polynomial vector
field $\ff = (f_1, \dots, f_{12})^T$ as follows:
\begin{align*}
  f_1 & =  \kf1 x_{5} -  \kr1 x_{1} -  \kf2 x_{1} +  \kr2 x_{12}\\
  f_2 & = \kf7 x_{11} -  \kr7 x_{2} -  \kf8 x_{2} +  \kr8 x_{9} + 
        \kf{10} x_{3} -  \kr{10} x_{2} -  \kf{15} x_{2} +  \kr{15}
        x_{10}\\
  f_3 & =  \kr4 x_{12} +  \kf6 x_{5} -  \kr6 x_{3} -  \kf{10} x_{3}
        +  \kr{10} x_{2} +  \kf3 x_{4} -  \kr3 x_{3} -  \kf4
        x_{3}\\
  f_4 & =  \kf5 x_{6} -  \kr5 x_{4} -  \kf9 x_{4} +  \kr9 x_{11} -
         \kf3 x_{4} +  \kr3 x_{3}\\
  f_5 & =  \kf0 x_{6} -  \kf6 x_{5} +  \kr6 x_{3} -  \kr0 x_{5} -
         \kf1 x_{5} +  \kr1 x_{1}\\
  f_6 & = - \kf0 x_{6} -  \kf5 x_{6} +  \kr5 x_{4} +  \kr0 x_{5}\\
  f_7 & =  \kf{13} x_{10} -  \kr{13} x_{7} +  \kf{16} x_{9} - 
        \kr{16} x_{7}\\ 
  f_8 & = - \kf{12} x_{8} +  \kr{12} x_{10} +  \kf{14} x_{11} - 
        \kr{14} x_{8}\\
  f_9 & =  \kf8 x_{2} -  \kr8 x_{9} +  \kf{11} x_{12} -  \kr{11}
        x_{9} -  \kf{16} x_{9} +  \kr{16} x_{7}\\
  f_{10} & =  \kf{12} x_{8} -  \kr{12} x_{10} -  \kf{13} x_{10} + 
        \kr{13} x_{7} +  \kf{15} x_{2} -  \kr{15} x_{10}\\
  f_{11} & = -\kf7 x_{11} +  \kr7 x_{2} +  \kf9 x_{4} -  \kr9 x_{11} -
            \kf{14} x_{11} +  \kr{14} x_{8}\\
  f_{12} & = - \kr4 x_{12} -  \kf{11} x_{12} +  \kr{11} x_{9} + \kf{2}
        x_{1} -  \kr2 x_{12} +  \kf4 x_{3}.\numberthis\label{eq:epspkin}
\end{align*}
In the presentation of the model in the BioModels repository, occurrences of
reaction rates $\kk$ are generally multiplied with the volume of a compartment
\ce{comp1}. This amounts in \eqref{eq:epspkin} to a corresponding factor for all
$\ff$, which would not affect our computations here and can be equivalently
dropped. It is noteworthy that our framework would allow to handle occurrences
of various different compartment volumes as extra parameters in $\kk$.

Our computations for this model did not finish within 24 hours, even when fixing
all forward reaction rates $\kf{i}$ to their values specified in the BioModels
repository. This is a bit surprising, because with regard to $|\kk|$, $|\xx|$,
and $|\ff|$, the problem is smaller than 5-site phosphorylation, which we
successfully computed in the previous section. Furthermore, $\ff = 0$ is a
system of parametric \emph{linear} equations. It seems that there is an immense
combinatorial explosion in the size of parametric coefficient polynomials caused
by iterated solving for certain variables and plugging in.

\section{Conclusions}\label{se:conclusions}

Geometric definitions of shifted toricity and toricity of a real steady state
variety $V$ require that $V \cap \Rsn$ forms a multiplicative coset or group,
respectively. We have proposed a formal framework, based on first-order logic
and real quantifier elimination, to test this in the presence of parameters.
Computational experiments succeeded on dynamics of reaction networks with up to
54 species and 30 parameters.

With all our computations on real-world networks here, we have found that the
coset property is independent of the choice of parameters. This result is
desirable from the viewpoint of chemical reaction theory, which postulates that
relevant properties of networks do not depend on reaction rates. Given the coset
property, the stronger group property holds only for degenerate choices of
parameters in the sense that they satisfy algebraic equations. In the context of
our framework, this is not too surprising. The equivalent conditions in the
parameters for the group property are obtained by plugging in $1$ for all
species concentrations in the defining equations of $V$. Our conclusion is that
the coset property without algebraic conditions on the parameters is the
relevant concept.

We have used above a strict notion of \emph{algebraic}, which excludes order
inequalities. Recall that we had advertised in the introduction that our
approach is capable of producing semi-algebraic conditions on the parameters,
which can include inequalities. Such inequalities come into existence during
quantifier elimination as sign conditions on discriminants of non-linear
polynomials. With the Triangle network in Sect.~\ref{se:triangle} they almost
made their way into the output but were removed in the last moment by quantifier
elimination-based simplification. One of them has been presented in
\eqref{eq:gamma2}. Beyond that, our computations did not produce any order
constraints on the parameters. It is an interesting question, maybe also for the
natural sciences, whether there is a systematic reason for their absence. A
positive answer would also support alternative purely algebraic approaches to
toricity, e.g., based on binomial ideals.

\subsection*{Acknowledgments}
This work has been supported by the interdisciplinary bilateral project
ANR-17-CE40-0036/DFG-391322026 SYMBIONT
\cite{BoulierFages:18a,BoulierFages:18b}. We are grateful to our project partner
Ovidiu Radulescu for helping us understand part of the biological background.

\providecommand{\noop}[1]{}


\end{document}